\documentstyle[12pt,fleqn]{article}

\def\beqa{\begin{eqnarray}}
\def\eeqa{\end{eqnarray}}
\def\beq{\begin{equation}}
\def\eeq{\end{equation}}

\def\hm{h^{-1}{\rm~Mpc}}
\def\vr{\mbox{\bf r}}

\let\gam=\gamma

\newcommand{\mincir}{\raise -2.truept\hbox{\rlap{\hbox{$\sim$}}\raise5.truept
\hbox{$<$}\ }}
\newcommand{\magcir}{\raise -2.truept\hbox{\rlap{\hbox{$\sim$}}\raise5.truept
\hbox{$>$}\ }}
\newcommand{\minmag}{\raise-2.truept\hbox{\rlap{\hbox{$<$}}
\raise 6.truept\hbox
{$>$}\ }}

\def\etal{{\it et al.}\ }

\def\rnat{ Nature }
\def\rapj{ Ap. J. }

\def\raaa{ Astron. Astrophys. }

\def\rmnras{ Mon. Not. R. Ast. Soc. }

\begin{document}
\thispagestyle{empty}
\begin{titlepage}
\vspace{.2in}
\title{{\hfill\normalsize Fermilab-Pub-94-043-A}\\
{\hfill\normalsize February 1994}\\
\vspace{.3in}
\bf Non-Gaussian statistics of pencil beam surveys}
\vspace{.4in}
\author{Luca Amendola}
\maketitle
\vspace{.1in}
\centerline{Osservatorio Astronomico di Roma}
\centerline{Viale del Parco Mellini, 84}
\centerline{00136 Rome - Italy}
\centerline{\it and}
\centerline{NASA/Fermilab Astrophysics Center
\footnote{Address until Oct. 1994; e-mail amendola@fnas06.fnal.gov}}
\centerline{Fermi National Accelerator Laboratory}
\centerline{PO Box 500}
\centerline{Batavia IL 60510 - USA}

\vspace{.2in}

\begin{abstract}
We study the effect of the non-Gaussian clustering of galaxies
 on the statistics of pencil beam surveys. We find that the higher order
moments of the galaxy distribution play an important  role in the
probability distribution for the power spectrum peaks. Taking into account
the observed values for the  kurtosis of galaxy
distribution we derive the general probability distribution for the power
spectrum modes in non-Gaussian models and show that the probability to
obtain

the 128$\hm$ periodicity found in pencil
beam surveys is raised by roughly one order of magnitude.
The non-Gaussianity of the galaxy distribution is however still

insufficient to explain the reported peak-to-noise ratio of the
periodicity, so that extra power on large scales
seems required.

\end{abstract}
\vspace{.8in}

{\bf Subject headings}: cosmology: large-scale structure of the Universe;
galaxies: clustering
\end{titlepage}

\section{Introduction}
\label{sec:intro}
The surprising discovery of a 128$\hm$ periodicity in the
distribution of galaxies (Broadhurst \etal 1989) has raised an
intense debate about the statistical significance of the signal
detected. The main question  is whether

the periodicity is consistent with the local observations or
is rather to be regarded as a new feature appearing only when very
large scales ($\gg 100 \hm$) are probed. In the original paper
by Broadhurst \etal (1989; BEKS) the statistical significance
of the
peak in the one-dimensional power spectrum was assessed making use
of an external estimator, i.e. adopting a model for the clustering of
galaxies. The clustering was assumed to be described by the
usual correlation function $\xi(r)=(r/r_0)^{-\gam}$ up to
the scale of 30$\hm$, without any correlation beyond this scale, and
without any higher order moment. As Szalay \etal (1991) pointed out,
however, external estimators are very model dependent. Even
slightly different assumptions, concerning e.g. selection functions or the
 parameters $r_0,\gam$, can result in dramatic variations of
statistical
significances. Indeed, Kaiser \& Peacock (1991), investigating

essentially the same dataset as BEKS, found that the noise level
was to be significatively raised, resulting in a much
higher probability to find
a peak as large as, or larger, the one at 128$\hm$, so as to
reconcile the standard model of galaxy clustering with the BEKS data.
Similarly, Luo \& Vishniac (1993) found that the redshift distortions
can alter the estimate of the noise level.

This seems to force one to use  internal estimators of the noise
level. This has been done
  by Szalay \etal (1991), who  showed then that the probability to find
a peak as high as the one in the BEKS  data, or higher, is $2.2\cdot 10^{-4}$,
matching their original estimate.
Luo \& Vishniac (1993) also confirmed the result that, while the
rest of
the power spectrum is consistent with the hypotheses of clustering and
Gaussianity, the single prominent spike at 128$\hm$ is not.

They also showed that even
 a delta-like feature in the tridimensional power
spectrum of the galaxy distribution can barely account for the BEKS spike.
 As we will show below, the
probability estimate on which these conclusions are based relies
 essentially on two hypotheses: $a)$ that
the spatial bins of the BEKS survey are uncorrelated, i.e. that the
clustering beyond 30$\hm$ is negligible, and $b)$ that the components of the
power spectrum  can be assumed, by virtue of the central limit
theorem, to be Gaussian distributed. The very fact that the probability
estimate based on these two hypotheses is as low as $2.2\cdot 10^{-4}$
points to the conclusion that one of the two, or both, are false.
This implies either that there is some

previously unknown, and theoretically unexpected, feature in the

tridimensional
power spectrum at large scale, or that it is the other hypothesis, the
Gaussianity,
to be abandoned.

The first possibility has been explored for instance in the Voronoi

simulations
(see e.g. Coles 1990, SubbaRao \& Szalay 1992), or in truncated
HDM models (Weiss \& Buchert 1993). Unlike the
precedent studies,
in this paper we consider in detail the latter way out.

The scheme of this paper is as follows.
 First, we derive the probability
distribution of the components of a one-dimensional
power spectrum in presence of higher order moments of the
spatial distribution. Second, we ask ourselves which is
the probability to find a  spike as high as, or higher than, the one in the
BEKS data in such non-Gaussian galaxy distribution. Finally,
adopting  the actual higher order moments found in local ($\le
100 \hm$) observations, we will show that the formal
probability for the
BEKS periodicity increases   roughly by an
order of magnitude. This, however, may still be insufficient to
explain the data.

Let us note that we will not question in any way the reliability of the
BEKS data or of their noise estimate. Rather, we derive our
conclusion only taking into due account the {\it already known} level
of non-Gaussianity in the galaxy distribution.

\section{Non-Gaussian pencil beam statistics}

The BEKS data consist in a set of counts along a survey
geometry that approximates a long, thin cylinder directed towards
the galactic poles. The galaxy positions are binned  in $N$ small cylinders
of radius $R=3\hm$ and radial length $30\hm$, out to $L/2\sim 1000\hm$
in both directions. The details of the survey
are given in the original paper (BEKS) and in Szalay \etal (1991).
 Let us denote the cell counts as $n_i$, with
$i=1,..N\approx 67$.  The
discrete Fourier transform of the dataset is
\beq\label{dft}
f_k={1\over P} \sum_{j=1}^{N} n_j \exp(i2\pi k r_j/L)\,,
\eeq
where $r_j=30 j \hm$ is the radial distance to the $j-th$ bin, and
$P=\sum n_j$ is the total number of galaxies
(396 in BEKS). The counts $n_j$ have mean $\hat n=P/N$ and variance
 $\sigma^2=<(n_j-\hat n)^2>$ as well as higher order irreducible moments
(or cumulants, or disconnected moments) $k_n$.

 The power spectrum is defined as
\beq
A_k=|f_k|^2\,.
\eeq
Let us define the quantity
\beq
a_k={\sum_j n_j \cos(2\pi k r_j/L)\over
\sigma \sqrt{N/2}}\,.
\eeq
Squaring $a_k$ we obtain
$
a_k^2=  ({\rm ~Re}f_k)^2/[
\sigma^2 (N/2P^2)]\,.
$
Likewise, we can define the  quantity $b_k = \sum_j n_j \sin(2\pi k r_j/L)/
\sigma \sqrt{N/2}$ and form the modulus
\beq
z\equiv a_k^2+b_k^2={A_k\over \sigma^2(N/2 P^2)}\,.
\eeq
The problem is now to find out the probability distribution
density (PDD)
of $A_k$ when we know the one for $n_j$.
First, however, we have to derive the PDD of $a_k$ and $b_k$.
They are constructed as a linear sum of $N$ independent variables (as
long as the various $n_j$ are uncorrelated), so by the central
limit theorem $a_k,b_k$ should tend to be Gaussian
distributed. However, since $N\sim 70$ is not really very large,
one should check if the higher order terms are significant. This is
indeed what will be shown to happen. We make use of the
so-called Edgeworth expansion (see e.g.  Cramer 1966, Abramovicz
\& Stegun 1972, whose notation we will follow), according to which
 the variable
\beq\label{form}
X={\sum_i (Y_i - m_i)\over (\sum_i \sigma_i^2)^{1/2}}
\eeq
(the sums run over $N$ terms)
where the $Y_i$ are independent random variables with mean
$m_i$ , variance $\sigma_i$ and $n$-th order cumulants $k_{n,i}$,
is distributed like a function $f(X)$ that can be expanded in

powers of $N^{-1/2}$
\beq\label{edge}
f(X)\sim G(X)\left[
1+{\gam_1\over 6N^{1/2}} He_3(X) + {\gam_2\over 24 N}He_4(X)+
{\gam_1^2\over 72 N}He_6(X) + O(N^{-3/2})\right]\,.
\eeq
Here, $G(X)$ is the normal distribution,
 $He_n$ is the Hermite polynomial of order $n$, and
$
\gam_1=(\sum_i k_{3i}/N)/(\sum_i
\sigma_i/N)^3\,,\quad \gam_2=(\sum_i k_{4i}/N)/(\sum_i \sigma_i/N)^4\,. $
The Edgeworth expansion has been used recently in astrophysics
by several authors
to quantify slight deviations from Gaussianity (Juszkiewicz \etal 1993,
and references therein).
Now we can notice that, as long as the counts
$n_j$ are uncorrelated,   the variables $a_k$ and $b_k$ are indeed  in the form
(\ref{form}),
where $Y_i=n_i \cos(2\pi k r_i/L)$

[or $Y_i=n_i \sin(2\pi k r_i/L)$ ] so we are allowed to

 apply the Edgeworth expansion.  To the order $1/N$,  the expansion
will  include the skewness
and  the kurtosis of the counts $n_i$.

However, the skewness sum $\sum k_{3i}$
 for the variables $Y_i$ vanishes due to the
oscillating term, so that $\gam_1=0$.

Let us estimate then the expansion coefficient
$\gam_2$ in our case.
 The higher order moments in the galaxy
counts have been calculated by several authors for  different surveys
(Saunders \etal 1991; Bouchet, Davis
\& Strauss 1992; Gazta\~naga 1992;
Loveday \etal 1992).
The general result is that, for  scales
which range  from some megaparsecs to more than 50 $\hm$,
 the dimensionless cumulants
$\mu_m=k_m/\hat n^m$ (for $m=2\,$, $~\mu_2=\sigma^2/\hat n^2$)

 obey the hierarchical scaling relation

\beq\label{sca}
\mu_m=S_m \mu_2^{m-1}\,,
\eeq
where $S_m$ are the scaling constants
(we have
checked that the shot-noise correction is negligible in our case).
  To the lowest order in the variance and for scales much larger
than the correlation length of the fluctuation field, the scaling relation
 is in reality a direct consequence of the Edgeworth
expansion (and can actually be derived by a much simpler
argument, see Amendola \& Borgani 1994).  Then we see that

\beq
\gam_2\approx (3/2) S_4 \mu_2\,.
\eeq
where the numerical factor is due to the sum over the sines and cosines
in $\sum_i k_{4,i}$.
The relations between the scaling
constants and the physics of the clustering process have
been investigated in several works, from the
book of Peebles (1980) to recent generalizations as in Bernardeau (1992).

 The value
of  $\mu_2=\sigma^2/\hat n^2$ can be expressed as a function of

the correlation
function (e.g. Peebles 1980), $\sigma^2=(\hat n+ \hat n^2\xi_0)$, where
$\xi_0=V^{-2}\int d^3r_1 d^3r_2 W(r_1)W(r_2) \xi(\vr_1-\vr_2)$ and
 where $W$ is the window function
corresponding to the BEKS cylindrical cells of volume $V$.
Since $\hat n\approx 6$ and $\xi_0$ is of order unity,
we can  approximate $\mu_2$ with $\xi_0$, so that

$\gam_2\approx (3/2) S_4 \xi_0$.
The value of $\gam_2$ will result to be  crucial.

Several uncertainties, however, prevent its exact estimate.
For $\xi_0$ we must rely on very local observations; we
may assume the value given in Szalay \etal (1991),

$\xi_0\approx 0.83$, or the one that we derive from
 Luo \& Vishniac (1993), $\xi_0\approx 1.24$, or similar values, depending
on models of the correlation function.
For $S_4$ one problem is that
we need the scaling constants for quite elongated
 cylindrical cells,  while the
observations have been carried out  mostly for  large spherical or cubic
cells.  We can find observational values from near unity to 30 or 40

(see e.g. the table in Gazta\~aga (1992)).  Further,
 Lahav \etal (1993) find that  $S_3$ and $S_4$, rather than being
constants, sharply increase
with the  rms density contrast $\delta$, and thus decrease with the
cell  volume,  when $\delta>1$ in CDM simulations.
We then absorb  the uncertainties of $S_4$  and
of $\xi_0$ in $\gam_2$ and explore numerically the
range $\gam_2\in (0-40)$.

Let us come back to the Edgeworth PDD $f(a_k)$ for $a_k$
(to the order $1/N$). The PDD for $y=a^2_k$ is then $P(y)=
f(a_k)(da_k/dy)=f(y^{1/2})/2y^{1/2}$, that is
\beq\label{edgechi}
P(y=a^2_k)\sim
g_1 P_1+{\gam_2\over 24 N}\left[g_5P_5-6g_3P_3+3g_1P_1\right] \equiv \sum_i c_i
P_i(y),
\eeq
 where $g_n\equiv 2^{n/2}\Gamma(n/2)$ and
$P_n(y)=g_n^{-1}y^{n/2-1}e^{-y/2}$ is the $\chi^2$ PDD with
$n$ degrees of freedom.

Now that we have the PDD for
$a_k^2, b_k^2$ we must find the distribution for $z=a^2_k+b^2_k$.
Let us denote with $\phi(t)$ the characteristic function (CF) of
a generic probability distribution $P(x)$, where
$~
\phi(t)=\int e^{itx} P(x) dx\,.
$
The general theorems about probability distributions say that the
CF of the sum of two variables is the product of the CF of the variables.
Furthermore, by linearity, we see that the CF of  $P=P_1+P_2$ is
$\phi(P_1)+\phi(P_2)$. We are to use these two properties to
derive the general distribution $P(A_k)$.
 First, we calculate the CF $\phi(P)$ for $P(y)$ given by (\ref{edgechi}),

 $P(y)=\sum_i c_i P_i$.
Denoting  the CF for the $\chi^2$ distribution $P_n$ as $\psi_n\equiv
(1-2it)^{-n/2}$, we have
\beq
\phi[z]=\phi[a^2_k]\phi[b^2_k]=\phi^2[y]=(\sum_i c_i \psi_i)^2\,,
\eeq
where the sum runs over all the $\chi^2$ PDD in the

expansion (\ref{edgechi}),

with the same $c_i$'s. Now, since $\psi_n\psi_m=\psi_{n+m}$,
we can see that the CF for the unknown distribution

$P(z)$ is a sum of $\chi^2$ CFs, so that the final result $P(z)$ is again a
sum of $\chi^2$ PDDs.

Before writing down the  result, we note that
$z=A_k/ [\sigma^2(N/2 P^2)]= 2 A_k/A_0$, where $A_0$ is the

noise level in the notation of BEKS.
It follows $A_0=\sigma^2 N/P^2=(\xi_0/N+ 1/P)$,
which gives an external estimate of the noise level. However, as already
remarked, the
estimate of $A_0$ by Szalay \etal (1991) is internal in that is
not based on a {\it a priori}
model for $\xi(r)$, but rather on fitting the observational distribution
function
for $A_k$ {\it at small amplitudes} with the exponential
$P(A_k)=(1/A_0)\exp (A_k/A_0)$,  as it should be in the purely
Gaussian case (or for $N\to\infty$).  The same internal
estimate applies here, since as we will see  the Gaussian and non-Gaussian
PDD are equivalent  at low amplitudes.

Finally, the normalized distribution function for
$A_k$   to the order $1/N$ is
\beq\label{final}
P(z=2A_k/A_0)=P_2+a(P_6-2P_4+P_2)
\eeq
where
$
a= \gam_2/ 4N.$

Eq. (\ref{final}) gives then  the general PDD for the power spectrum

amplitudes relative
to a set of pencil beam counts with  scaling coefficient   $S_4$.

When $S_4=0$ we return to the exponential distribution $P(z)=P_2$
on which the calculation of BEKS, and of all the other works on the subject,
 was based.
We can see from $P(z)$ why the higher order terms are important. Since the
peak-to-noise
ratio $X\equiv A_k/A_0$ found by BEKS is very large, $X_{BEKS}=11.8$, the terms
containing higher order $\chi^2$ functions will
dominate over the $P_2$ term when integrated to give the cumulative
probability,
even if the constant $a$ are small, i.e. even if $N$ is large.
Actually, for any given $N$ there is a value  $z_c$ such as  the higher order
terms
dominate over the lower orders in the integral
$\int_{z_c}^{+\infty} P(z)dz$ . This is
a consequence of the  fact that,
 while the convergence of any
distribution $f(X)$ to the normal one for $N\to\infty$
is ensured by the central limit theorem, the convergence itself need not
be uniform. The fractional difference between the cumulative distribution of
$f(X)$ and the
one relative to
a normal distribution can be arbitrarily large  for large deviations from the
mean.

We can now  directly compare the PDD (\ref{final}) with the power spectrum
coefficients found by BEKS. We use the tabulated values provided
by Luo \& Vishniac (1993), binned in peak-to-noise intervals of $0.5$.
We plot in Fig. 1 the cumulative function of the BEKS coefficients

versus peak-to-noise ratio
(a point at abscissa $x$
represents the fraction of values of $A_k$ in the BEKS
data with peak-to-noise ratio larger than $x$)
 and compare this with our theoretical cumulative
function
\beq
F(X)=\int_{2X}^{+\infty} P(z)dz\,,
\eeq
where $X=A_k/A_0=z/2$ is the peak-to-noise ratio.
The functions plotted are for the Gaussian case ($S_4=0$),
and for three possible values of the  constant $\gam_2$: from bottom to top,

$\gam_2=5, 20, 40$. It is clear that as $\gam_2$ increases, the

observed distribution becomes more and more
consistent with the non-Gaussian behavior,
 except for the last point, the 128$\hm$
spike, which

appears still far away from its expected
frequency value. However, we can estimate now
the probability to have $X_{BEKS}=11.8$ or higher in one
of the $\sim 30$ $k$-bins to which BEKS assigned the data
(see Szalay \etal 1991 for a detailed exposition) and compare with

the very unlikely value $2.2\cdot 10^{-4}$ originally found for $\gam_2=0$.
 The non-Gaussian result is
\beq\label{ngr}
P(>11.8)\approx 30 F(11.8)=0.001-0.005\,,
\eeq
for the range $\gam_2=10-40$.

The inclusion of non-Gaussianity pushed the probability to obtain
the BEKS spike by about one order of magnitudes,  without any need
to invoke non standard features in the galaxy distribution.
The result (\ref{ngr}) states that the BEKS spike should
occur roughly in  0.1-0.5\% of the cases if the very
large scale galaxy distribution has to be consistent with the
local observations of variance and kurtosis. If further data
do not reduce the peak significativity, our result

indicate that is very difficult for the non-Gaussianity alone to
explain the observations.

In Fig. 2 we display the behavior of $P(>11.8)$ {\it vs.}
 $\gam_2$ .
   Only for very large values
of $\gam_2$, and thence of $S_4$ or of $\xi_0$, the BEKS spike
approaches the 3$\sigma$ level. In other words, if the BEKS periodicity
is strenghtened by further data, we will be forced to assume values of
$S_4$ or $\xi_0$ larger than local observations would require, and/or
to discard the assumption that the spatial bins
are uncorrelated.  On the other
hand, a value of $X_{BEKS}$ smaller by even a ten percent would
result in quite higher values of $P(>11.8)$, as shown by the dot-dashed
curve in Fig. 2.

\section{Conclusions}

We have shown that the higher order moments of the galaxy
clustering play a not negligible  role in assessing the significance
level of peaks in one-dimensional power spectra. The scaling
constant $S_4$ and the correlation average $\xi_0$ on
the spatial bin in a pencil beam survey are combined in
the crucial parameter $\gam_2$. Assuming that the
spatial bins are uncorrelated, the question raised by
the remarkable periodicity discovered by Broadhurst \etal (1989)
in the very large scale galaxy clustering can then be expressed in
the following way: are the values of $S_4$ and of $\xi_0$

determined by local observations compatible with the clustering
of galaxies at the very deep scales probed by the pencil beam?
To give an answer, we have to determine the probability distribution for
the power spectrum amplitudes of  a non-Gaussian field sampled in
spatial bins. We find by means of the Edgeworth expansion that
the BEKS most prominent spike around 128$\hm$ has a probability
of roughly   $1-5 \cdot 10^{-3}$ for acceptable
 values of $\gam_2$, to be compared
with the value $2.2\cdot 10^{-4}$ obtained by Szalay \etal (1991)
neglecting the kurtosis correction.
This result seems to show that non-Gaussianity alone
 cannot be responsible for the BEKS
periodicity, unless further observations will allow very
large values of $S_4$ or $\xi_0$ or will decrease the peak-to-noise
ratio of the 128$\hm$ peak. As the data stand, we should conclude that
the spatial bins cannot be assumed uncorrelated.

Indeed, this is what
the large coherent structures reported in deep surveys

seems to require.
 We also compared the peak occurrences of
the full spectrum of BEKS and found  it in a good agreement with
our non-Gaussian probability distribution if large
values for $\gam_2$ are allowed. This raises the

interesting possibility that
further pencil beam data can be employed to measure  the
parameter $\gam_2$, i.e.  the product $S_4\xi_0$, down to very
deep distances.

Let us conclude by  the remark
  that   is not unlikely that
further higher order terms in the Edgeworth expansion are significant.
Further terms  would require however the knowledge of

new scaling coefficients, like $S_6$ and so on, which are not
available.
Here we confined ourselves to the order $1/N$ for simplicity,
 with the
aim to show how the non-Gaussian properties of the galaxy clustering
have a strong effect on the peak probability estimate. In this sense,
our calculation gives only a lower bound on the probability estimate.

\vspace{.2in}
\centerline{\bf Acknowledgments}
I thank Stephane Colombi for useful discussions. This work has been
supported by DOE and NASA under grant NAGW-2381. I also
acknowledge   CNR (Italy) for financial support.

\newpage
\centerline{\bf References}
\vspace{.2in}
\frenchspacing
\noindent
  Amendola L. \& Borgani S. 1994, \rmnras 266, 203\\
  Bernardeau F. 1992 \rapj 392, 1\\
  Bouchet F., Davis M.,  \& Strauss M. 1992, in Proc. of the 2nd DAEC Meeting
on the

 Distribution of Matter in the Universe, eds. G.A. Mamon, \& D. Gerbal,

287  \\
  Bouchet F. R., \& Hernquist L. 1992, \rapj 25, 400  \\
  Broadhurst T.J., Ellis R.S., Koo D.C. \& Szalay A. 1990 \rnat
343, 726 \\
  Coles P. 1990 \rnat 346, 446 \\
  Coles P. \&  Frenk C.S. 1991, \rmnras, 253, 727 \\
  Cramer H. 1966, Mathematical  Methods of Statistics (Princeton:
Princeton Univ. Press)\\
  Gazta\~naga E. 1992, \rapj, 398, L17 \\
  Juszkiewicz R., \& Bouchet F.R. 1992, in Proc. of the 2nd DAEC Meeting on

the Distribution

 of Matter in the Universe, eds. G.A. Mamon, \& D. Gerbal,

301 \\
Juszkiewicz R., Weinberg D. H., Amsterdamski P.,
Chodorowski M. \& Bouchet F. 1993,

 preprint (IANSS-AST 93/50)\\
  Kaiser N. \& Peacock J.A. 1991, \rapj 379 482 \\
  Lahav O., Itoh M, Inagaki S. \& Suto Y. 1993, \rapj 402, 387 \\
  Loveday J., Efstathiou G., Peterson B. A. \& Maddox
S.J. 1992, \rapj, 400, L43 \\
  Luo S. \& Vishniac E.T. 1993, \rapj 415, 450. \\
  Peebles P.J.E. 1980, The Large-Scale Structure of the Universe (Princeton:
Princeton Univ.

Press) \\
  Saunders W., Frenk C., Rowan-Robinson M., Efstathiou

G., Lawrence A., Kaiser N., Ellis

 R., Crawford J., Xia X.-Y., Parry
I. 1991, \rnat, 349, 32  \\
  SubbaRao M. \& Szalay A. 1992, \rapj 391, 483 \\
  Szalay A.S., Ellis R.S., Koo D.C. \& Broadhurst T.J. 1991, in
Proc. After the First

Three Minutes, ed. S. Holt, C.
Bennett \& V. Trimble (New York: AIP) \\
  Weiss A. \& Buchert T. 1993, \raaa 274,1

\newpage
\centerline{\bf Figure Caption}
{\bf Fig. 1.}\\
Cumulative probability distribution for the Gaussian model (dashed straight
line),
for the non-Gaussian models with   $\gam_2=5,20,40$ (solid lines, bottom to
top), and for the BEKS data (filled squares).  The vertical long-dashed line
marks the peak-to-noise ratio of BEKS, $X=11.8$.
\vspace{.3in}

{\bf Fig. 2.}\\
Probability to find a peak as high as, or higher than, the 128$\protect\hm$
peak of BEKS
as a function of the crucial parameter $\gam_2$.  The
long-dashed curve is for a value of $X$ ten percent lesser than
$X_{BEKS}=11.8$.
The horizontal dashed lines show the rejection levels of 99.7 \%.

\end{document}